\documentstyle[eqsecnum,aps,twocolumn,epsf]{revtex}
\newcommand{\p}{\partial}

\begin{document}
\draft
\preprint{LPT-Orsay 01/32}
\title{Kinetic equilibration in heavy ion collisions:\\
The role of elastic processes}
\author{Julien Serreau}
\address{Institut f\"ur Theoretische Physik der Universit\"at Heidelberg,\\
Philosophenweg 16, D-69120 Heidelberg, Germany}
\author{Dominique Schiff}
\address{Laboratoire de Physique Th\'eorique,\\
 B\^{a}timent 210, Universit\'e Paris-Sud, 
91405 Orsay, France}
\date{\today}
\maketitle
\begin{abstract}
 We study the kinetic equilibration of gluons produced in the 
 very early stages of a high energy heavy ion collision in a
 ``self-consistent'' relaxation time approximation. We compare 
 two scenarios describing the initial state of the gluon system, 
 namely the saturation and the minijet scenarios, both at RHIC and 
 LHC energies. We argue that, in order to characterize kinetic
 equilibration, it is relevant to test the isotropy of various 
 observables. As a consequence, we find in particular that in both 
 scenarios elastic processes are not sufficient for the system to 
 reach kinetic equilibrium at RHIC energies. More generally, we show
 that, contrary to what is often assumed in the literature, elastic
 collisions alone are not sufficient to rapidly achieve kinetic 
 equilibration. Because of longitudinal expansion at early times, 
 the actual equilibration time is at least of the order of a few fermis.
\end{abstract}

\narrowtext

\section{Introduction}

The main interest of studying very high energy heavy-ion collisions
is the possibility that a quark-gluon plasma (QGP) be formed. It is widely 
believed that a large amount of real partons (essentially gluons) with 
transverse momentum $p_t \sim 1-2$~GeV is freed in the very early stages 
of such collisions, carrying most of the  produced transverse 
energy~\cite{minijet,Mueller}. While rapidly diluted by longitudinal 
expansion, this parton gas tends to locally equilibrate via mutual 
interactions between its constituents. When the mean free path of a 
parton becomes too large, the description in terms of partonic degrees of 
freedom ceases to be valid; one is instead left with an expanding gas of 
hadrons. Whether the dense parton gas reaches a state of kinetic equilibrium 
before it hadronizes is a question of great interest for the interpretation 
of the data at RHIC and LHC. So far, most of the calculations concerning 
experimental signatures of the QGP rely on the assumption that the system 
is at least kinetically equilibrated.
\par
It was commonly assumed for a long time (see {\it e.g.}~\cite{Biro}) 
that shortly after the nuclei collide, kinetic equilibrium is 
rapidly achieved on time scales $\lesssim 1$~fm (see {\it e.g.}~\cite{BMPR}). 
It was however shown that, due to the effect of expansion, at least elastic
collisions may not be effective enough~\cite{Wong,2to3,HeisWang}. An important 
question is also to characterize the initial state of the parton system 
formed just after the collision~\cite{Mueller,HeisWang}. The purpose 
of this paper is to reexamine the role of elastic collisions in the 
process of kinetic equilibration, assuming that, already at early times, 
a local Boltzmann equation may be written for the partonic phase space 
distribution. We specifically compare within a simple relaxation time 
approximation the case where initial conditions are given by the saturation 
scenario, which may already be valid at RHIC, to the case where the initial
gluons are produced incoherently in the perturbative QCD regime, the so-called 
minijet scenario. We consider only the pure gluon case and $2 \rightarrow 2$
elastic collisions in the limit of small-angle scatterings. By calculating 
the time-dependent relaxation time in a self consistent way, we are able to 
obtain various observables as functions of time, thus probing kinetic 
equilibration. The method is described in sections~\ref{model} to~\ref{solve}. 
In section~\ref{result}, we present our results. Comparing with Ref.~\cite{Raju},
where the exact solution is worked out in the saturation scenario, one can 
assess the validity of our approach, which appears to be surprisingly good. 
We then apply this method to the minijet scenario, considered in 
Ref.~\cite{Dumitru}, and we consistently take care of conservation laws. 
We argue that, to characterize equilibration, a reliable criterium is to 
test the isotropy of various observables. As a consequence we find in 
particular that for both initial conditions the system does not reach
kinetic equilibrium at RHIC energies. We discuss the reliability 
of our description and argue in particular that, due to the fragility 
of the weak coupling approximation, it appears difficult to obtain 
definite conclusions at LHC energies. In any case, the kinetic 
equilibration time is an order of magnitude bigger than the typical 
$1$~fm estimate usually assumed. The hypothesis that elastic collisions 
are sufficient to rapidly achieve kinetic equilibration, often made 
in the literature~\cite{Biro} is thus shown to be wrong, as already 
suggested in~\cite{Wong}. Similar conclusions have been obtained in a
slightly different context by the authors of Ref.~\cite{Gyulassy}.

\section{The Boltzmann equation in the small angle approximation}
\label{model}

Shortly after the gluons have been produced, the occupation number becomes 
small enough so that one can treat the further evolution by means of a 
Boltzmann equation describing the time evolution of the local
distribution $f(\vec p,\vec x,t)$ for on-shell particles\footnote{One can 
   argue that the requirement for using a local Boltzmann equation for on-shell 
   quanta is that the occupation number $f$ be $\lesssim 1/\alpha_S$, 
   where $\alpha_S$ is the strong coupling constant~\cite{Muellerprivate}.}. 
When the occupation number becomes smaller than unity, one can make the further
approximation of a gas of classical particles. Here, we consider only the 
$gg \rightarrow gg$ elastic processes in the collision term, which we shall 
treat in the leading small-angle scattering approximation. Assuming 
one-dimensional expansion in the early stages after the collision and 
longitudinal boost invariance in the central region in the center of mass 
frame, the Boltzmann equation can be written in terms of a derivative at
constant $p_z t$~\cite{Baym}:
\begin{equation}
\label{BE}
 \left. \p_t f(\vec p,t) \right|_{p_z t} = {\mathcal C} (\vec p,t) \, .
\end{equation}\noindent
Defining the moments $N_s=\langle p^s \rangle = \int_{\vec p} \, p^s \, f(\vec p,t)$,
with the notation $\int_{\vec p} \, \equiv \, 2(N_c^2-1) \int d^3p/(2\pi)^3$,
one obtains for the collision term $\mathcal C$, in the limit of small-angle 
elastic scatterings between classical particles ($f \ll 1$), to logarithmic
accuracy~\cite{Mueller,Landau},
\begin{equation}
\label{Coll}
 {\mathcal C} (\vec p,t) = 
 {\mathcal L} \, N_0 \nabla_p^2 f(\vec p,t) + 
 2 {\mathcal L} \, N_{-1} \vec\nabla_p \left[\vec v \cdot f(\vec p,t)\right]  \, ,
\end{equation}\noindent
where $\vec v = \vec p/p$, and where 
$ {\mathcal L} = \pi \alpha_S^2 \frac{N_c^2}{N_c^2-1} \ln ( 1/\chi_{min}^2 )$, 
$\chi_{min}$ being the minimum scattering angle, whose value is determined
by the physics of Debye screening. We shall come back to this later. 
\par
In the present paper, we will solve the Boltzmann equation in a relaxation 
time approximation, that is we will solve the equations for moments of the 
distribution $f$, taking into account however the microscopic dynamical 
information contained in (\ref{Coll}).

\section{The ``self-consistent'' relaxation\\time approximation}

The Relaxation Time Approximation (RTA) simply consists in replacing the
effect of the collisions in the Boltzmann equation (\ref{BE}) by an 
exponential-like relaxation towards a local ``equilibrium'' 
distribution\footnote{The distribution $f_{eq}$ has the meaning of 
   a local equilibrium distribution only when the system has reached the
   hydrodynamic regime, where $\lambda (t)$ and $t^{1/3} T(t)$ are 
   constants~\cite{Bjorken}}:
${\mathcal C}\equiv - (f-f_{eq})/\theta$, with $f_{eq}(\vec p,t) = \lambda (t) 
\, \exp (-p/T(t))$.
We assume that the relaxation time $\theta$ does not depend on $\vec p$.
This has to be viewed as a mean-field like approximation : the different
modes are effectively decoupled and their evolution is governed by an
effective relaxation time which contains the microscopic information
and is to be computed self-consistently. Indeed, strictly speaking this 
equation has no solution and one has to take it in a weaker sense, that is 
at the level of moments of the partonic distribution:
\begin{equation}
\label{momeq}
 \int_{\vec p} \, m(\vec p,t) \, {\mathcal C} (\vec p,t) =  
 - \frac{\langle m \rangle (t) - \langle m \rangle_{eq} (t)}{\theta_m (t)} \, ,
\end{equation}\noindent
where $m(\vec p,t)$ is an arbitrary function, $\langle m \rangle_{(eq)} = 
\int_{\vec p} \, m(\vec p,t) \, f_{(eq)}(\vec p,t)$, and $\theta_m (t)$ is
the associated relaxation time, different choices for the function $m$ leading
to different relaxation times. 
\par
Let us be more precise on this last point: it is clear 
that the RTA, being a one-time-scale ansatz, cannot fully reproduce the 
evolution of the whole ensemble of modes. A particular choice for the 
function $m$ picks up a particular momentum scale in the momentum distribution 
$f$, whose contribution to the moment $\langle m \rangle$ dominates. The 
relaxation toward isotropy of this part of the distribution is characterized 
by the time scale $\theta_m$. So the appropriate choice for the moment function 
$m$ depends on the momentum scale characterizing the physics one wants to study.
\par
In this paper we are concerned with the question of kinetic equilibration, 
that is with the isotropy of ``thermodynamic'' quantities, like longitudinal
and transverse pressures $P_L (t) = \langle p_z^2/p \rangle$ and
$P_T (t) = \langle p_\perp^2/p \rangle$, where $p_\perp^2 = (p_x^2 + p_y^2)/2$.
The particles we are interested in have energy of the order of the average 
energy per particle $\bar \epsilon (t) = \epsilon (t)/n(t)$,
($n=N_0$ and $\epsilon = N_1$ are the particle number
and energy densities per unit volume respectively).
A pertinent choice leading to simple equations is\footnote{The choices 
   $\langle m \rangle = P_L$ or $\langle m \rangle = P_T$ give the same 
   results as those presented below. The choice $\langle m \rangle = \epsilon$ 
   leads to the equation of energy conservation (see Eq.~(\ref{engy})) which 
   contains no information about collisions.} 
$m(\vec p) = (p_z^2 - p_\perp^2)/p$ or equivalently, 
$\langle m \rangle = P_L - P_T$. Inserting the Landau-Mueller 
collision term (\ref{Coll}) in (\ref{momeq}) and introducing the 
notation\footnote{For example, $P_L=N_1^z$, $P_T=N_1^\perp$.}:
$N_s^z = \langle p_z^2 \, p^{s-2} \rangle$ and 
$N_s^\perp = \langle p_\perp^2 \, p^{s-2} \rangle$, one obtains the 
following equation for the relaxation time:
\begin{equation}
\label{momeq1}
  \frac{N_1^z - N_1^\perp}{\theta} = 
  4 \, {\mathcal L} \, N_0 \, ( N_{-1}^z - N_{-1}^\perp ) +
  2 \, {\mathcal L} \, N_{-1} \, ( N_0^z - N_0^\perp ) \, .
\end{equation}\noindent
\par
The two other parameters of the relaxation time ansatz, $\lambda (t)$ and 
$T(t)$, are computed using conservation laws: conservation of energy, and
of particle number in elastic collisions\footnote{The ``fugacity''
   parameter $\lambda (t)$ is needed to enforce particle number 
   conservation in elastic collisions.}. 
In the RTA, these give the relations
\begin{eqnarray}
\label{engy}
 \epsilon (t) & = & \epsilon_{eq} (t) = 
 6 \frac{N_c^2-1}{\pi^2} \, \lambda(t) \, T^4(t) \, , \\
\label{numb}
 n(t) & = & n_{eq} (t) = 
 2 \frac{N_c^2-1}{\pi^2} \, \lambda(t) \, T^3(t) \, .
\end{eqnarray}
Note that the particle number conservation law implies that the number
density exactly falls like $t^{-1}$: $t \, n(t)=t_0 \, n(t_0)$, with $t_0$ the 
time when our description begins.

\section{Solving the equations}
\label{solve}

The solution of the Boltzmann equation in the RTA is given by~\cite{Baym}
\begin{eqnarray}
 f(\vec p_t,p_z,t)&=&f_0(\vec p_t,p_z\frac{t}{t_0}) \, \mbox{e}^{-x(t)} \nonumber\\
 && + \int_{t_0}^t dt' \, \frac{\mbox{e}^{x(t')-x(t)}}{\theta(t')} \,
 f_{eq} (\vec p_t,p_z\frac{t}{t'},t') \, ,
\label{baym}
\end{eqnarray}\noindent
where $x(t) = \int_{t_0}^t dt' \, \theta^{-1}(t')$, $f_0 (\vec p)$ is the initial
distribution, and where $\lambda(t)$, $T(t)$ and $\theta^{-1}(t)$ are computed 
at each time with the help of Eqs.~(\ref{momeq1})-(\ref{numb}). We can use
Eq.~(\ref{baym}) to give an integral expression for all the quantities appearing 
in these equations. In the present case, every phase space integral can be 
computed analytically, thus considerably simplifying the numerical task. 
One typically obtains a formula of the type
\begin{eqnarray}
 M (t) & = &  M(t_0) \, {\mathcal F}_M^{(0)} (t_0/t) \, \mbox{e}^{-x} \nonumber\\
 && \, \, \, + \int_{t_0}^t dt' \, \frac{\mbox{e}^{x'-x}}{\theta'} \,
 {\mathcal F}_M^{(eq)} (t'/t) \, M_{eq} (t') \, ,
\label{MBaym}
\end{eqnarray}\noindent
where $x \equiv x(t)$, $x' \equiv x(t')$, $\theta' \equiv \theta (t')$ and 
$M \equiv N_s$, $N_s^z$ or $N_s^\perp$ ($s=-1,0,1$), and where the corresponding
functions ${\mathcal F}_M^{(0)}$ and ${\mathcal F}_M^{(eq)}$ are calculable. 
The function ${\mathcal F}_M^{(0)}$ depends on the form of the initial 
distribution. The ``equilibrium'' moments $M_{eq}$, computed with $f_{eq}$, 
only depend on $\lambda$ and $T$ (see Eqs.~(\ref{Amoment0}) and
(\ref{Amomentz}) in the appendix).
\par
We solve the system of equations (\ref{momeq1})-(\ref{numb}) and (\ref{MBaym})
by using a method similar to that described in Ref.~\cite{Dumitru}: suppose
we know the parameters $\lambda$, $T$ and $\theta^{-1}$ for every time $t\le t_1$.
We want to compute their values at time $t_1 + \delta t$. We use a first guess
(for example the values at time $t_1$) to compute a first estimation $M^{(1)}$ of 
the moments $M(t_1 + \delta t)$, using Eq.~(\ref{MBaym}). We then compute a 
second estimation of our three parameters with the help of 
Eqs.~(\ref{momeq1})-(\ref{numb}), with which we obtain a second estimation 
$M^{(2)}$ of $M(t_1 + \delta t)$, again using Eq.~(\ref{MBaym}). We repeat
these steps until convergence of our three parameters to some fixed accuracy.
We have checked the validity of this method by solving Eq.~(\ref{MBaym}) in some
exactly solvable cases, for example by computing $n=N_0$ at each time-step with
the help of Eq.~(\ref{MBaym}), with 
${\mathcal F}_n^{(0)} (a) = {\mathcal F}_n^{(eq)} (a) = a$.

\section{Initial conditions and results}
\label{result}

In this section, we apply the above formalism to the study 
of kinetic equilibration of the gluon gas formed during the very
early stages of the collision. We study two different scenarios
for the initial condition, namely the saturation and minijet scenarios.
The corresponding initial distributions are taken from Ref.~\cite{Raju}
for the former case (see also~\cite{Mueller}) and from Ref.~\cite{Dumitru}
for the later. 
\par
To proceed further, one has to give a prescription for regularizing
the logarithmic singularity ${\mathcal L}$ appearing in the collision 
term (Eq.~(\ref{Coll})), in the small angle approximation.

\subsection{Screening}
\label{screening}

The logarithmically divergent integral appearing in the small 
scattering angle collision integral is physically regulated by 
screening effects in the gluon medium~\cite{Raju,Landau,Debye}. 
The minimum scattering angle is given by the relation 
$m_D^2 = \underline p^2 \, \chi_{min}^2$, where $m_D$ is the Debye 
mass, an expression of which can be obtained in terms of the distribution 
function in the linear response approximation~\cite{Debye}, 
and where $\underline p$ is the typical momentum of particles in the medium.
We shall see below that for early times, because of the expansion, the
particles in the central region have essentially zero longitudinal 
momentum\footnote{This is explicit in the initial distribution corresponding
   to the saturation scenario (see Eq.~(\ref{insat})). In the minijet scenario,
   although the initial distribution is isotropic in momentum space, it becomes
   rapidly peaked around $p_z=0$ (see Fig.~\ref{fig_relaxjet}).}.
In this situation, the exchanged gluon is essentially transverse and the
relevant screening mass is the transverse mass~\cite{Raju}
$$
 m_T^2 (t) = \frac{\alpha_S N_c}{\pi^2} \, \int \frac{d^3p}{p} \, f(\vec p,t) =
 \frac{4 \pi \alpha_S N_c}{N_c^2-1} \, N_{-1} (t) \, ,
$$
and we write $L \equiv \ln (1/\chi_{min}^2) = 
\ln (\overline{\langle p_t^2 \rangle}/m_T^2)$, where $p_t^2=p_x^2 + p_y^2$ and
where $\overline{\langle ... \rangle}=\langle ... \rangle/n$ denotes the 
average per particle. Although the above choice for $L$ is motivated by the
highly anisotropic form of the distribution at early times, we shall use it
also when the system approaches kinetic equilibrium\footnote{For an 
   isotropic distribution, a more appropriate choice would be 
   $L_{iso}=\ln (\overline{\langle p^2 \rangle}/m_D^2)$. In this case 
   however $m_D^2 = 2 m_T^2$, $\langle p_t^2 \rangle=
   2 \, \langle p^2 \rangle /3$, and $L-L_{iso}= \ln 4/3 \simeq 0.3$.}.    
It should be noted here that the validity of the logarithmic approximation
is marginal because of the fact that the coupling constant is not very small
compared to one (in particular, one has typically $\alpha_S N_c \simeq 1$, 
see below)\footnote{For more details on this point, see~\cite{these}.}. 

\subsection{Saturation scenario}

In the saturation scenario one assumes that the produced transverse
energy per unit rapidity is dominated by small-x gluons having
a transverse momentum $p_t \sim Q_s$, the saturation momentum~\cite{Mueller}. 
These gluons are essentially freed during the
collision, after a time $t_i \sim 1/Q_s$ which is also the time needed for
gluons in different units of rapidity to physically separate from each
other. After a time $t_i$, gluons populating the central region
in the center of mass frame of the collision have essentially no longitudinal
momentum ($p_z t_i \sim p_z / Q_s \ll 1$). Following Mueller~\cite{Mueller}, 
one writes for the occupation number in the central region at $t_i$:
$$ 
 f_{sat} (\vec p,t_i) = \frac{c}{\alpha_S N_c} \, \delta (p_z t_i) \,
 \Theta (Q_s^2 - p_t^2) \, ,
$$
where $c \sim 1$ is a numerical constant, which has been computed numerically
using a classical field approximation in Ref.~\cite{KraVenc}. For an
$SU(2)$ gauge theory, $c=1.3$. The approximation of classical
particles makes sense when the occupation number $f \lesssim 1$ so that,
following the authors of Ref~\cite{Raju}, we take, as initial time,
$t_0 = \frac{c}{\alpha_S N_c} \, t_i$ where one has, assuming free 
streaming between $t_i$ and $t_0$,
\begin{equation}
\label{insat}
 f_0 (\vec p) = f_{sat} (\vec p,t_0) = 
 \delta (p_z t_i) \, \Theta (Q_s^2 - p_t^2) \, .
\end{equation}\noindent
The values of the parameters corresponding to RHIC and LHC energies given 
in~\cite{Raju} are summarized in Tab.~\ref{tab_sat}. In the following we 
take $\alpha_S=0.3$, $N_c=3$. We evolve the system until the particle number 
density becomes lower than $1$ per fm$^3$, when the description in terms 
of partonic degrees of freedom becomes meaningless. The corresponding time 
$t_{max}$ is evaluated using the conservation of the particle number: 
$t \, n(t) = t_0 \, n(t_0)$ (see Tab.~\ref{tab_sat}). This rough estimate 
serves only to give an upper limit for our description. In particular one 
obtains $t_{max} \simeq 30$~fm at LHC, which is too long for the 
approximation of longitudinally boost invariant geometry to be valid. 
Nevertheless, we present our results for $t \le t_{max}$ in order to show 
the behavior of the system.
\par
\begin{figure}
\epsfxsize=3.5in \centerline{ \epsfbox{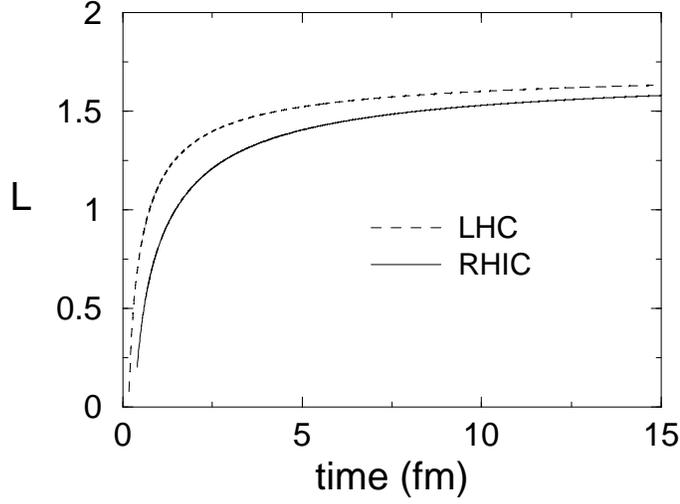}}
\caption{\small The logarithm $L$ (see section~\ref{screening}) as a function 
of time (in fm) in the saturation scenario at RHIC and LHC.} 
\label{fig_logsat}
\end{figure}

Fig.~\ref{fig_logsat} shows the logarithm $L$ as a function of time. At early 
times, $L\ll 1$ whatever the way one chooses to implement screening. As time 
goes on, although the situation becomes better, the logarithmic approximation 
remains marginal.
\par
\begin{figure}
\epsfxsize=3.5in \centerline{ \epsfbox{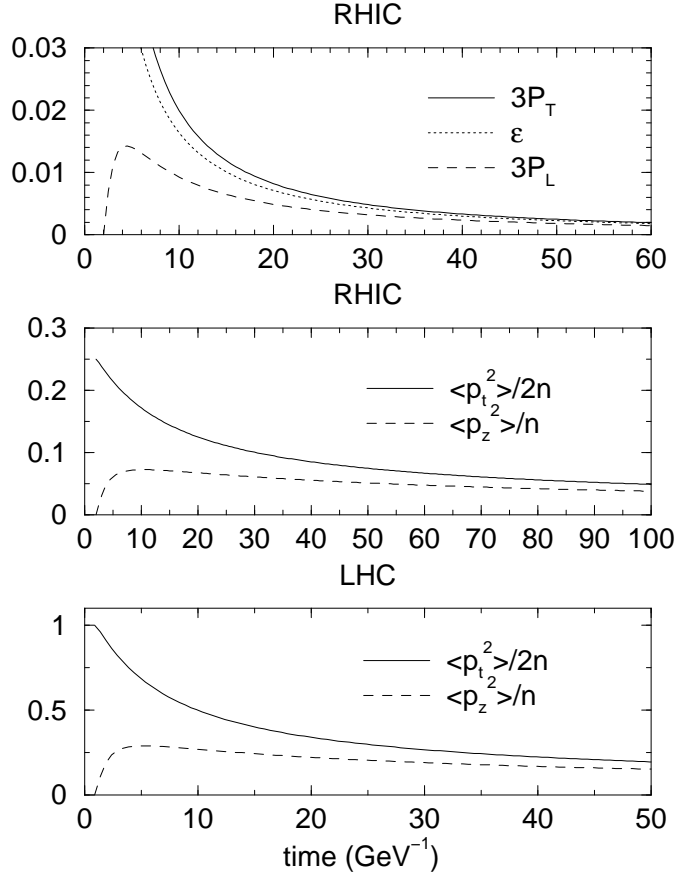}}
\caption{\small Various observables as functions of time in the 
saturation scenario. The upper panel shows the time evolution of 
the energy density and of the longitudinal and transverse pressures 
(times $3$) at RHIC. The second and third panels show the evolution
of the average longitudinal and transverse squared momenta per particle
$\overline{\langle p_z^2 \rangle}$ and $\overline{\langle p_\perp^2 \rangle}$
at RHIC and LHC respectively. Comparing with the curves obtained in [8],
one observes a semi-quantitative agreement (in the above figure, everything 
is expressed in units of GeV, as in [8]).}
\label{fig_raju}
\end{figure}

We first test our relaxation time approximation by comparing our
results with those of Ref.~\cite{Raju}. We show in Fig.~\ref{fig_raju}
the time evolution of the energy density $\epsilon$ and the longitudinal
and transverse pressures $P_L$ and $P_T$ at RHIC, as well as those of
the mean longitudinal and transverse squared momentum per particle
$\overline{\langle p_z^2 \rangle}$ and $\overline{\langle p_\perp^2 \rangle}$, 
both for RHIC and LHC energies. These are to be compared with Figs. 7 and 9
of~\cite{Raju} (we choose the units so as to make the comparison easiest):
one observes a semi-quantitative agreement with the exact solution, which
is remarkably good in view of the simplicity of our approach.
\par
\begin{figure}
\epsfxsize=5.5in \centerline{ \epsfbox{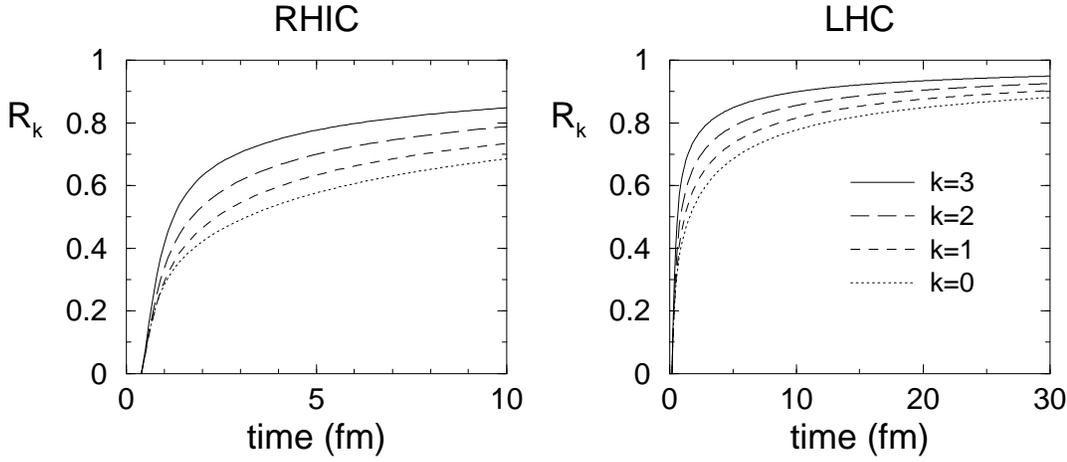}}
\caption{\small Time evolution of the ratios $R_k$, defined in Eq.~(\ref{ratio})
in the saturation scenario at RHIC (left) and LHC (right). The quantities $1-R_k$
give a measure of the anisotropy of the microscopic distribution $f(\vec p,t)$
for different momentum scales, the larger $k$ the lower the probed momentum 
scale. These curves are obtained with $\alpha_S =0.3$ and the logarithm
$L=\ln (\overline{\langle p_t^2 \rangle}/m_T^2)$ (see section~\ref{screening}).}
\label{fig_relaxsat}
\end{figure}
\begin{figure}
\epsfxsize=5.5in \centerline{ \epsfbox{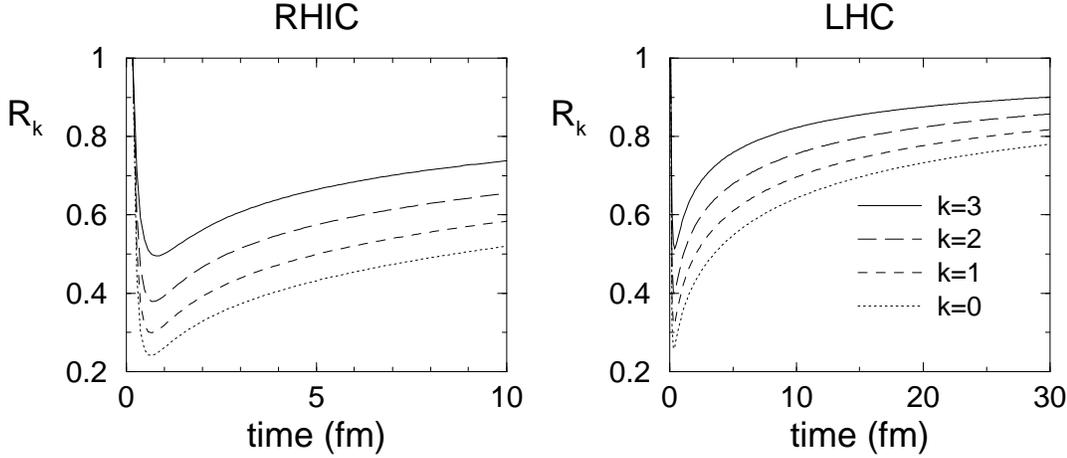}}
\caption{\small Time evolution of the ratios $R_k$, defined in Eq.~(\ref{ratio})
in the minijet scenario at RHIC (left) and LHC (right). These curves are 
obtained with $\alpha_S =0.3$ and the logarithm 
$L=\ln (\overline{\langle p_t^2 \rangle}/m_T^2)$ (see section~\ref{screening}).}
\label{fig_relaxjet}
\end{figure}

We now come to the study of kinetic equilibration of the gluon gas.
To this end, we measure the anisotropy of the distribution by means
of the ratios
\begin{equation}
\label{ratio}
 R_k = \frac{\langle p_z^2 / p^k \rangle}{\langle p_\perp^2 / p^k} =
 \frac{N_z^k}{N_\perp^k} \, ,
\end{equation}\noindent
which should approach $1$ as the system approaches kinetic equilibrium.
We stress here that the criterium we use to characterize kinetic equilibrium
(the isotropy of the distribution) is quite different from that used
in~\cite{Raju}. As a consequence, although we have essentially similar
results, we arrive at quite different conclusions. Indeed, the time evolution
of $R_k$ for $k=0,\ldots ,3$, at RHIC and LHC energies, is shown in
Fig.~\ref{fig_relaxsat}, where one can see that the system is far from
being isotropic even at $t \sim 10$~fm at RHIC. The situation looks 
better at LHC, where $R_k \simeq 0.8$ for $t \simeq 10$~fm. 
It is to be noted that in this scenario, the only scale is $Q_s$ and 
consequently the results for RHIC and LHC are essentially the same when 
everything is expressed in units of $Q_s$.

\subsection{Minijet scenario}

In the minijet scenario, the initial gluons are produced by means of hard 
and semi-hard collisions between the partons of the incident nuclei, and one 
can compute the initial energy and particle number density in perturbative
QCD~\cite{minijet}. The initial gluon multiplicity is dominated by the softest
of these hard gluons, having an energy $p_0 \sim 1-2$~GeV, which are produced
on a typical time $t_0\ \sim 1/p_0$. Following the author of Ref.~\cite{Dumitru}, 
we parametrize the initial state by a Boltzmann distribution. In the central 
region, we have
\begin{equation}
\label{injet}
 f_0 (\vec p) = f_{jet} (\vec p,t_0=1/p_0) = 
 \lambda_{jet} \, \mbox{e}^{-p/T_{jet}} \, ,
\end{equation}\noindent
where the parameters $\lambda_{jet} = \lambda (t_0)$ and $T_{jet} = T(t_0)$ 
are determined from the initial energy and particle number densities: 
$\epsilon_{jet} = \epsilon (t_0) = \epsilon_{eq} (t_0)$ and 
$n_{jet}=n(t_0)=n_{eq} (t_0)$. We recall the values of the parameters used
in~\cite{Dumitru} for RHIC and LHC in Tab.~\ref{tab_jet}. As in the previous 
section, we use $\alpha_S = 0.3$, $N_c=3$. Concerning the validity of the
logarithmic approximation, the situation is more confortable here: in the
time interval we considered, we have $1.5 \lesssim L \lesssim 5$. 
\par
Fig.~\ref{fig_relaxjet} shows the time evolution of the ratio $R_k$ defined
in (\ref{ratio}) for both RHIC and LHC energies. As expected, one observes 
a regime a free streaming at very early times, until the effect of collisions 
begins to set in, that is for time $t-t_0 \ll \theta(t_0)$. Concerning the 
degree of isotropy the system reaches, the situation is even worth than in 
the saturation scenario: the system is always far from isotropy at RHIC; at 
LHC, the system is still quite anisotropic ($R_k \lesssim 0.8$) until 
$t \sim 30$~fm. Here again, we note that the only scale of the description
is $T_{jet}$, and the results for RHIC and LHC are essentially the same when
expressed in units of $T_{jet}$.
\par
The above conclusions are to be compared with those of Ref.~\cite{Dumitru}, 
where the authors obtain relatively short ``equilibration times'', of the 
order of $4-5$~fm, both at RHIC and LHC. However, although only elastic 
collisions between gluons are considered, the total number of particles
is not conserved in~\cite{Dumitru}, and in fact grows with time (the number
density $n$ falls slower than $1/t$). This means in particular that 
collisions between particles are anomalously frequent and consequently
the system equilibrates anomalously fast. The similarity of equilibration
times obtained in~\cite{Dumitru} at RHIC and LHC is due to the fact that 
the authors work with a running coupling constant which grows with time,
as the typical momentum of the particles decreases. The case of a running 
coupling constant is examined in next section. We shall see that our 
previous conclusions remain unchanged.

\subsection{Discussion}

Let us first remark that the model used here to describe
the evolution of the gluon gas can only give qualitative 
information. Here we use it to study the question: are the 
elastic collisions efficient enough to achieve kinetic 
equilibrium, and if so on which time scale? In order to 
obtain a reliable answer, let us study the sensitivity of 
the above results to the details of the description,
for example the value of $\alpha_S$ or the prescription used to cut-off
the logarithmic divergence in the small-angle limit.

\subsubsection{Running coupling constant}

It is interesting to consider the case where the coupling constant is
allowed to run, growing with time as the typical energy of the particles
in the medium decreases. As mentionned above, this should accelerate the
equilibration process. Here we shall compute the coupling constant at each
time as a function of the mean energy per particle:
$\alpha_S (t) \equiv \alpha_S (\mu=\bar\epsilon=3T)$, with
\begin{equation}
\label{running}
 \alpha_S (\mu) = \alpha_S (M_Z) \, 
 \frac{\ln (M_Z/\Lambda_{QCD})}{\ln (\mu/\Lambda_{QCD})} \, ,
\end{equation}\noindent
where $M_Z \simeq 90$~GeV the mass of the $Z^0$ boson, $\alpha_S (M_Z)=0.1$,
and where we take $\Lambda_{QCD} \simeq 200$~MeV.
\par
\begin{figure}
\epsfxsize=5.5in \centerline{ \epsfbox{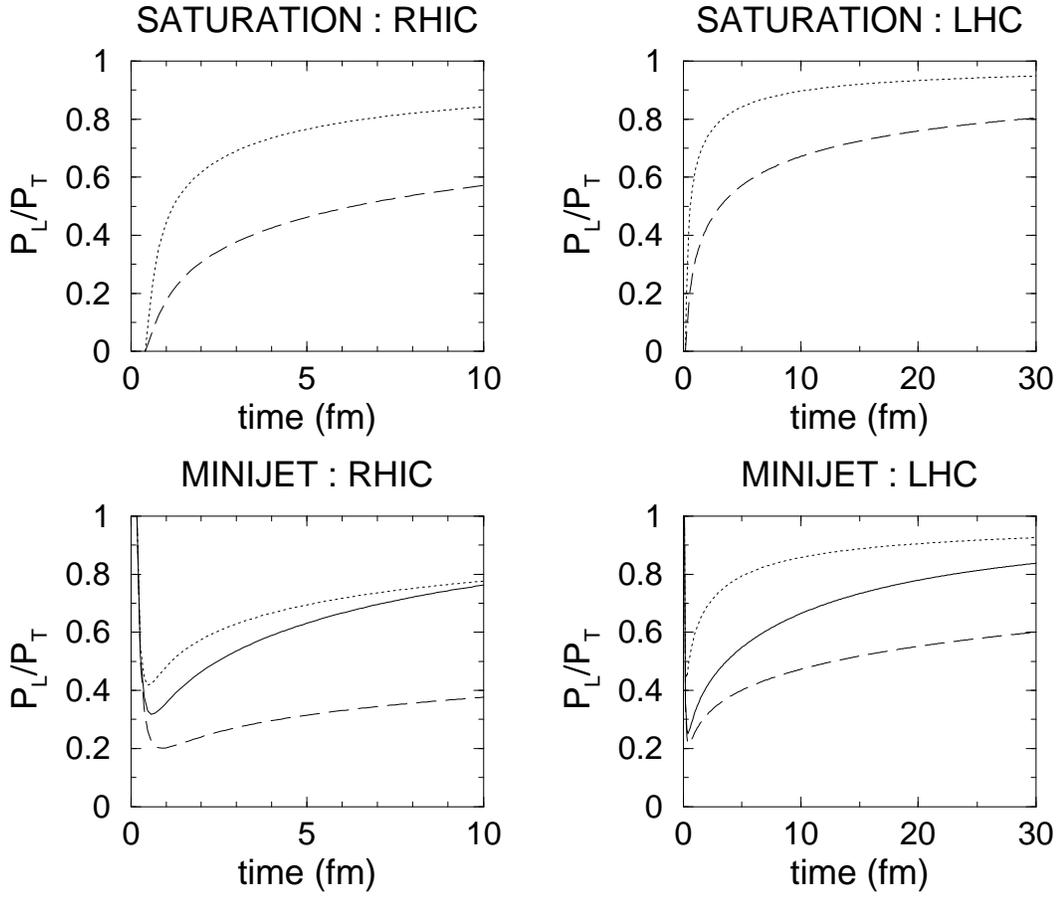}}
\caption{\small Sensitivity of the results (here for the time evolution of the
ratio $R_1=P_L/P_T$) with respect to the details of the description in the 
saturation (upper panels) and minijet (lower panels) scenarios at RHIC 
(left panels) and LHC (right panels) energies. Compared to the results shown 
in Figs.~\ref{fig_relaxsat} and~\ref{fig_relaxjet}, the dotted curves are obtained
by replacing the logarithm $L \rightarrow 2L$ and the dashed curves by replacing
$L \rightarrow L/2$. The solid curves in the minijet scenario correspond to 
the case where the coupling constant runs with time (see Eq.~(\ref{running})).} 
\label{fig_rob}
\end{figure}

In the minijet scenario, at LHC, we have $0.2<\alpha_S (t)<0.4$ on the
time interval we considered, and we obtain very similar results to the
case where $\alpha_S=0.3$. At RHIC, $0.3<\alpha_S (t)<0.5$ and, although 
the situation is better when compared to the fixed $\alpha_S$ case, the
system still does not reach kinetic equilibrium. The corresponding curves for
the ratio $R_1=P_L/P_T$ are shown in Fig.~\ref{fig_rob}.
\par
In the saturation scenario, the mean energy per particle is smaller than
in the previous case and the coupling constant becomes rapidly $\gtrsim 0.5$.
and the results have no physical significance. This is to be viewed as a 
manifestation of the fragility of the weak coupling approximation.

\subsubsection{Robustness of the results}

In order to estimate roughly the accuracy of our results, we give two
estimates, simply replacing $L \rightarrow 2L$ and $L \rightarrow L/2$ 
in the collision term, keeping $\alpha_S=0.3$.
The corresponding curves for the ratio $R_1=P_L/P_T$ are shown in 
Figs.~\ref{fig_rob} for both scenarios, at RHIC and LHC energies.
In the minijet scenario we also show the curves corresponding to the
case where the coupling constant depends on time, which are seen to 
be contained within our ``uncertainty bands''. One observes that the 
qualitative conclusions of previous sections are unchanged: in both
scenarios, elastic collisions are not sufficient to drive the system
towards kinetic equilibrium at RHIC energies. At LHC, kinetic equilibrium
may be reached in the saturation scenario even for times reasonnably short
for the approximation of longitudinal expansion to have a sense, say 
$t \sim 10$~fm. However, it appears difficult to draw any firm conclusion 
because of the uncertainties of our qualitative description. In the minijet 
scenario at LHC energy the situation is even worst and one cannot draw 
definite conclusions concerning kinetic equilibration even for very long 
times. What one can say however is that, in both scenarios, the assumption 
that kinetic equilibration is rapidly achieved by means of elastic collisions 
only, on time scales $\lesssim 1$~fm, is not reliable at RHIC energy and
questionnable at LHC energy, where the actual equilibration time is at 
least of the order of a few fermis.

\section{Conclusion}

We have studied the kinetic equilibration of the gluon gas initially produced
in very high energy nuclear collisions, by considering $2 \rightarrow 2$ 
small angle elastic scatterings only. Two different scenarios have been assumed
for the production of such gluons: the saturation and the minijet
scenarios. By using a simple ``self-consistent'' relaxation time approximation, 
we are able to reproduce semi-quantitative features of the exact solution 
of the Boltzmann equation. By measuring the anisotropy of different observables 
as a function of time, we can follow the system toward kinetic equilibrium.
Our results show that elastic collisions are not as efficient as usually
believed to achieve kinetic equilibrium. This contradicts a widely used 
assumption.
\par
This does not means that kinetic equilibrium is not achieved in a real 
collision. Indeed, concerning the physical question of kinetic equilibration,
the present study is only preliminary as one should include other contributions
in the collision integral. In particular, inelastic branching processes have
recently been proposed as playing a very important role in this 
respect~\cite{2to3}.

\acknowledgments

We acknowledge very useful discussions with R. Baier, A. Krzywicki and 
A. H. Mueller. LPT is laboratoire associé au CNRS - URA00063.

\appendix
\section*{Moments of the distribution}

We give here the explicit form of the equilibrium moments $M_{eq}$ and
of the functions ${\mathcal F}_M^{(0,eq)}$ appearing in Eq.~(\ref{MBaym}), for the
different moments $M$ used in this paper. These are of three different
sort: ``isotropic'', as $N_s=\langle p^s \rangle$, ``longitudinal'',
as $N_s^z=\langle p_z^2 \, p^{s-2} \rangle$, and ``transverse'', 
$N_\perp^z=\langle p_\perp^2 \, p^{s-2} \rangle$, with the obvious
relation: $N_s = N_s^z + 2 N_s^\perp$. Introducing the notation
$$
 \left\{ A || B \right\} =
 \left\{
 \begin{array}{rl}
 A & \mbox{if } f_0 (\vec p) = f_{eq} (\vec p,t_0) \\
 B & \mbox{if } f_0 (\vec p) = \delta (p_z) \, g(p_\perp) \, ,
 \end{array}
 \right.
$$
one easily gets, using Eq.~(\ref{baym}), 
\begin{eqnarray}
 t \, \mbox{e}^{x} \, N_s (t) & = & t_0 \, N_s (t_0) \, 
 \left\{ h_s (t_0/t) || 1 \right\} \nonumber\\ 
 && + \int_{t_0}^t \frac{dt'}{\theta'} \,
 t' \, \mbox{e}^{x'} \, N_s^{eq} (t') \, h_s (t'/t) \, ,
\label{Amoment0}
\end{eqnarray}\noindent
and
\begin{eqnarray}
 t^3 \, \mbox{e}^{x} \, N_s^z (t) & = & t_0^3 \, N_s (t_0) \, 
 \left\{ h_s^z (t_0/t) || 0 \right\} \nonumber\\
 && + \int_{t_0}^t \frac{dt'}{\theta'} \,
 (t')^3 \, \mbox{e}^{x'} \, N_s^{eq} (t') \, h_s^z (t'/t) \, ,
\label{Amomentz}
\end{eqnarray}\noindent
where
$$
 N_s^{eq} (t) = \int_{\vec p} \, p^s \, f_{eq} (\vec p,t) =
 (s+2)! \, \frac{N_c^2-1}{\pi^2} \, \lambda (t) \, T^{s+3} (t) \, ,
$$
and where
\begin{eqnarray*}
 h_s (a) & = & \int_0^1 dx \, \left[ 1 - (1-a^2) x^2 \right]^{s/2} \, ,\\ 
 h_s^z (a) & = & \int_0^1 dx \, x^2 \, \left[ 1 - (1-a^2) x^2 \right]^{s/2-1} \, .
\end{eqnarray*}\noindent
Eqs.~(\ref{Amoment0})-(\ref{Amomentz}) can be written in the generic form of 
Eq.~(\ref{MBaym}), using the following definitions: 
${\mathcal F}_{N_s}^{eq} (a) = a \, h_s (a)$ and 
${\mathcal F}_{N_s^z}^{eq} (a) = 3 a^3 \, h_s^z (a)$.
\par
Writing $A=\sqrt{1-a^2}$, one obtains:
\begin{center}
\begin{tabular}{llrllr}
 & $\displaystyle  h_{-1} (a) = \frac{\arcsin A}{A}$ & , &
 & $\displaystyle h_1 (a) = 
       \frac{a + h_{-1} (a)}{2}$ , & \\
 &&&&& \\
 & $\displaystyle  h_{-2} (a) = \frac{\arctan A}{A}$ & , &
 & $\displaystyle h_2 (a) = \frac{2+a^2}{3}$ . & \\
\end{tabular} 
\end{center}
The functions $h_s^z$ can be deduced from the previous ones by using
the obvious relation
$$
 h_s^z (a) = \frac{2}{s} \, \frac{d}{d a^2} h_s (a) \, .
$$ 
For the particular case $s=0$, one obtains
$$
 h_0 (a) = 1 \, \, \, , \, \, \, 
 h_0^z (a) = \frac{1}{A^2} \, \left( \frac{\arctan A}{A} - 1 \right) \, \, .
$$

\newpage
\onecolumn

\mediumtext
\begin{table}
\begin{tabular}{c|ccccc}
\small{\bf SATURATION} & $Q_s$ (GeV) & $t_0$ (fm) & $n(t_0)$ (fm$^{-3}$) & 
 $\epsilon (t_0)$  (GeV/fm$^{3}$) & $t_{max}$ (fm) \\
\hline
RHIC &  1. & 0.4  &  18.1 & 12.0 & $\sim 10$ \\
\hline
LHC  &  2. & 0.18 & 163.4 & 217.9 & $\sim 30$ \\
\end{tabular}
\caption{Values of the parameters $Q_s$ and $t_0$ characterizing the initial 
state in the saturation scenario at RHIC and LHC energies.
The energy and particle number densities are computed from the initial 
distribution Eq.~(\ref{insat}). The time $t_{max}$ is the time after which 
the particle density becomes less than $1$/fm$^3$ and is evaluated 
using particle number conservation: $t \, n(t) = t_0 \, n(t_0)$.}
\label{tab_sat}
\end{table}

\begin{table}
\begin{tabular}{c|cccccc}
\small{\bf MINIJET} & $t_0$ (fm) & $n_{jet}$ (fm$^{-3}$) & $\epsilon_{jet}$
(GeV/fm$^{3}$) &  $\lambda_{jet}$ & $T_{jet}$ (GeV) & $t_{max}$ (fm) \\
\hline
RHIC & 0.18 & 34.3 & 56.0 & 1.0 & 0.535 & $\sim 10$ \\
\hline
LHC  & 0.09 & 321.6 & 1110.0 & 1.0 & 1.13 & $\sim 30$ \\
\end{tabular}
\caption{Values of the energy and particle number densities characterizing 
the state of the gluon system at the initial time $t_0$ in the minijet 
scenario, at RHIC and LHC energies.
The values of the parameters $\lambda_{jet}$  and $T_{jet}$ characterizing the
assumed initial distribution Eq.~(\ref{injet}) are obtained from 
Eqs.~(\ref{engy})-(\ref{numb}) at $t=t_0$. The time $t_{max}$ is the time 
after which the particle density becomes less than $1$/fm$^3$.}
\label{tab_jet}
\end{table}

\end{document}